# History of ARIES: A premier research institute in the area of observational sciences

Ram Sagar

Indian Institute of Astrophysics, Bangalore, 560034 and
Aryabhatta Research Institute of Observational Sciences, Nainital, 263001;
Email for correspondence: ram_sagar0@yahoo.co.in

## Abstract

The Aryabhatta Research Institute of Observational Sciences (ARIES), a premier autonomous research institute under the Department of Science and Technology, Government of India has a legacy of about seven decades with contributions made in the field of observational sciences namely atmospheric and astrophysics. The Survey of India used a location at ARIES, determined with an accuracy of better than 10 meters on a world datum through institute's participation in a global network of Earth artificial satellites imaging during late 1950's. Taking advantage of its high-altitude location, ARIES, for the first time, provided valuable input for climate change studies by long term characterization of physical and chemical properties of aerosols and trace gases in the central Himalayan regions. In astrophysical sciences, the institute has contributed precise and sometime unique observations of the celestial bodies leading to a number of discoveries. With the installation of the 3.6-m Devasthal optical telescope in the year 2015, India became the only Asian country to join those few nations of the world who are hosting 4-m class optical telescopes. This telescope, having advantage of geographical location, is well-suited for multi-wavelength observations and for sub-arc-second resolution imaging of the celestial objects including follow-up of the GMRT, AstroSat and gravitational-wave sources.

## 1 Introduction

The Aryabhatta Research Institute of Observational Sciences (ARIES) is a premier autonomous research institute under the Department of Science and Technology (DST), Government of India. Its head quarter is situated at Manora Peak (longitude = 79° 27′ E; latitude = 29°22′ N; altitude = 1950m), a few km from scenic town of Nainital in the central Himalayan region (Fig. 1). Through a cabinet approval of the Government of India, dated January 7, 2004, the 50 years old state Observatory was rechristened as ARIES on March 22, 2004 (Sagar, 2006). The institute name starting with Aryabhatta (spelled as Aryabhata in some literatures) was given as a tribute to one of the greatest ancient modern Indian astronomers. Incidentally, the Institute's acronym name ARIES, also signifies the zodiacal sign of the Sun at its two historically significant epochs separated by 50 years. First one is the birth of the organization on April 20, 1954 as Uttar Pradesh State Observatory (UPSO) under the Uttar Pradesh (UP) state government (Sinvhal, 2006) and the second one is the formation of ARIES under the Department of Science and Technology, Government of India, on March 22, 2004 (Sagar, 2006).

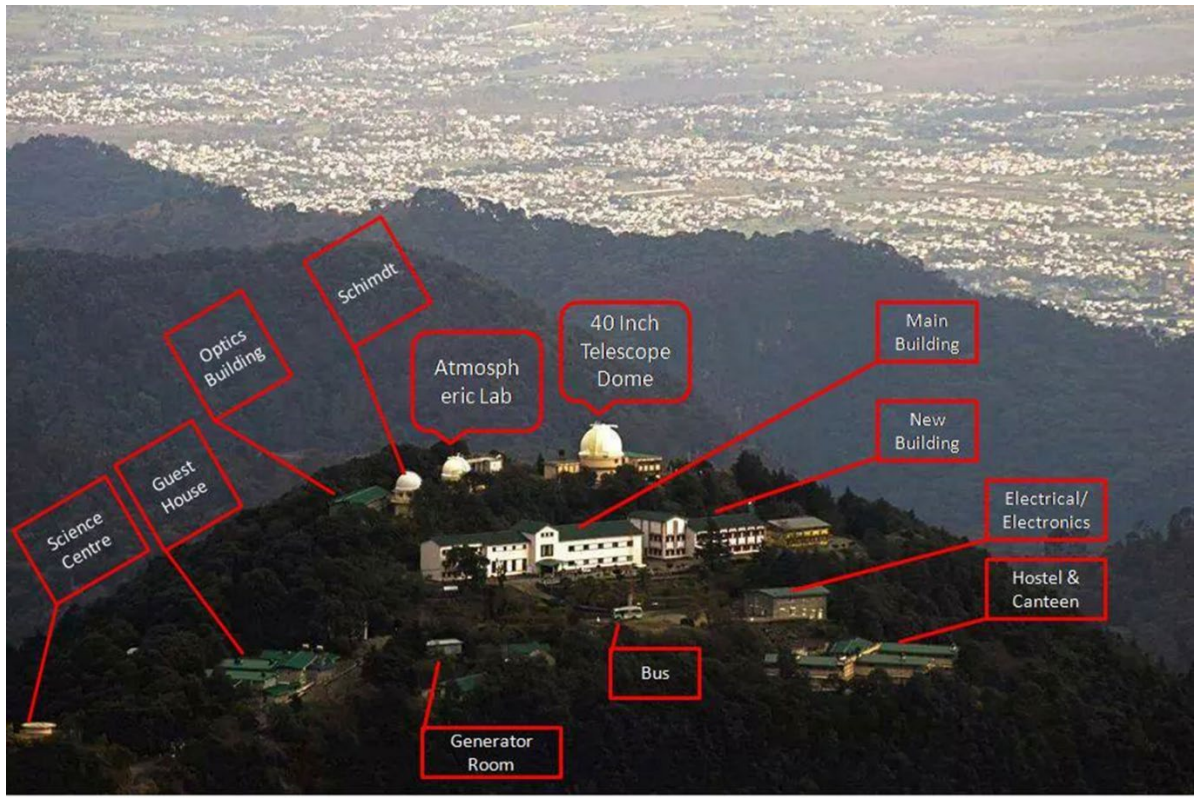

**Fig. 1** An aerial view of the topography and various buildings located around Manora peak top, main campus of the ARIES. The laboratory for atmospheric science is located between the domes of Schmidt and 104-cm (40 inch) telescopes, just above the main office building. Other main buildings seen are Science centre, Ashwini Guest house, Optics, Electrical/Electronic and Hostel.

After independence of India in the year 1947, many academic and research institutions were established by both centre and state governments. Idea of starting an astronomical observatory in the State of UP was germinated by Dr. Sampurnanand, the then Education minister and later the Chief Minister of the UP state. He, being a scholarly statesman, was interested in nurturing the science of Astronomy–the mother of a number of other branches in fundamental sciences. With the efforts of Dr Sampurnanand and Dr Awadesh Narayan Singh, a Professor of Mathematics at

Lucknow University, the UPSO was formed on April 20, 1954. At that time, it had the unique distinction of being the only pure science research institution in the field of Astronomy & Astrophysics funded by a state government in India. Further details on this matter can be found elsewhere (Sinvhal, 2006). Dr A.N. Singh was appointed the honorary Director of the Observatory and Dr S.D. Sinvhal was the first staff to join as Assistant Astronomer on April 20, 1954. The Observatory started its scientific activities in the holy city of Benaras (now Varanasi) at the Government Sanskrit College (now Sampurnanand Sanskrit Vishwavidyalaya). In July 1954, during the course of his inspection tour to the Observatory, Dr Singh unfortunately expired at Varanasi due to a sudden heart attack. Dr M.K. Vainu Bappu, then a young (~ 27 years old) PhD from Harvard School of Astronomy, United States of America (USA), joined the UPSO in Varanasi on November 1, l954 as its head (then called Chief Astronomer). He, being a motivated and enthusiastic person, started thinking of making this institution an international centre of astronomical research. Within a few months, he convinced the authorities of the UP government to shift this Observatory from dusty and hazy location of Varanasi to a place well suited for astronomical observations. A detailed description of this event is given by Sinvhal (2006). In November 1955, the Observatory was shifted over from Varanasi to a small cottage at Devi Lodge, halfway up to Snow View from the Lake Bridge at Nainital. Finally, in November, 1961, it moved to the present location of ARIES at Manora peak. The boundaries of ARIES are defined by the road to Nainital. Presently, it has 32.38 hectares of land at Manora peak and another 5.0426 hectares of land at Devasthal, situated at about 60 Km east by road from Manora peak.

In March 1960, Dr Bappu moved to the Kodaikainal Observatory as its Director and Dr S. D. Sinvhal took over the charge of the UPSO Director in April 1960. He held the position till October 1978 and again from June 1981 till his retirement in May 1982. Dr. M. C. Pande worked as officiating Director from November 1978 till May 1981 and then Director from June 1982 till his retirement in March 1995. In July 1996, author took over the charge of Director, and he continued till the formation of ARIES on 22 March 2004. Fig. 2 displays photographs of germinator of the UPSO along with all five Directors who led the institution under both UP and Uttaranchal state governments.

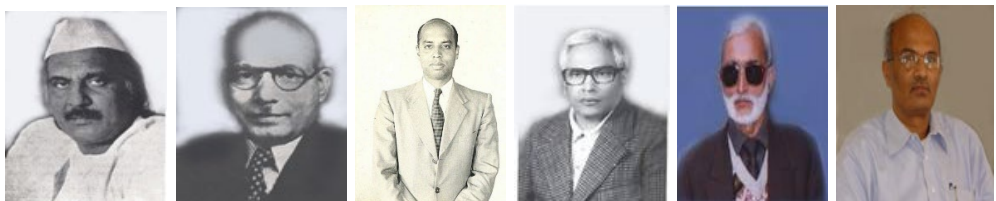

**Fig. 2** From left to right: Dr Sampurnanand, Dr A.N. Singh, Dr M.K. Vainu Bappu, Dr S.D. Sinvhal, Dr M.C. Pande and Prof. Ram Sagar.

This article describes history of the ARIES. The author had privilege of serving this institution for an extended period (from 1973 to 1979 as a scientist and from 1996 to 2014 as Director) and therefore is able to provide authentic information on the subject matter. It is well known that in

India, most of the pure science research institutions are nurtured mainly by the government. The change is the law of nature and the same can be seen in the history of ARIES as well. It was nurtured by both state and central governments for different durations. For over 46 years, UP State government nurtured the Observatory. Due to geographical division of the UP state on November 9, 2000, the Observatory became part of the Uttaranchal state. During reorganisation process of the newly formed state, the state Observatory was converted into ARIES on March 22, 2004 through a central government Cabinet decision dated January 7, 2004 (Sagar, 2006). The historical details of moving the institution from state government to the Department of Science and Technology under Government of India are mentioned in the section 2. The section 3 provides the history of observational facilities and scientific results obtained at ARIES in the field of observational sciences. The history of research publications is given in section 4, while last section 5 provides summary and conclusions.

## 2 Background of formation of ARIES under DST

During initial formative years of the UPSO, a number of optical telescopes and cameras were commissioned at Manora Peak, Nainital as observational facilities in the field of astronomy and astrophysics. However, research and academic activities of the institution being part of pure science, it was not in the priority list of the state government as reflected in the meetings held with authorities of the UP government coming mainly from Indian Administrative Services (IAS). For example, during summer of 1999, a high-level meeting was arranged in the Lucknow office of then UP Chief Secretary, Mr. Yogendra Narain, IAS regarding a large telescope project in collaboration with Tata Institute of Fundamental Research (TIFR), Mumbai. Other officers present in the meeting were Prof. V.S. Ramamurthy, then Secretary, DST, central government; Prof. S. S. Jha, then Director, TIFR and IAS Secretaries of finance, planning and science departments of the UP government. As Director, UPSO, the author was coordinating the meeting. For maintenance and operating the telescope after its successful installation, finance Secretary suggested to take night charges from the users. Vehemently opposing the idea, Prof. Ramamurthy stated that in scientific world, such practices are not followed. When I visited office of the finance Secretary in Lucknow for his signature on minutes of the meeting, he asked me to explain reasons why his idea of nightly charges was not appreciated while you must know that science department has least priority in the state government set up and you will face financial difficulties while operating and maintaining the telescope after installation with funding from the state government. I justified statements of Prof. Ramamurthy by citing examples of international practices followed in projects like Hubble space telescope etc. He was finally convinced and helped the institution later on.

In light of the above, after ~ 3 decades of its formation, the functioning of the Observatory became administratively and financially handicapped and academic activities initiated during that period could not be completed (Sanwal et al., 2018). The inflows of funds reduced significantly due to internal systems and procedures prevalent and applicable to government departments which were not conducive to an academic research institution. The institution was inhibited from proactively establishing linkages with similar national and international institutions to the required extent. The growth of the institution, consequently, became far below its potential. Prof. Jayant V Narlikar (Fig. 3), then the Director, Inter University Centre for Astronomy and Astrophysics (IUCAA), Pune, was the first to raise these issues in his letter dated April 28, 1998 addressed to the then Union Minister of Science & Technology, Prof. (Dr.) Murli Manohar Joshi. In the letter, he suggested that the institution should be taken over by the Government of India. Despite considering the letter positively, Prof. Joshi was of the view that the Government of India would not like to interfere in the affairs of a state government and necessary steps in this direction can be initiated by the center only if such a request comes from the UP government. The letter was, therefore, sent to the UP government for its consideration but nothing happened. However, this letter was taken seriously by the Uttaranchal government and it initiated action in December 2000, within months of its formation. In fact, during reorganization of the newly formed state, its authorities thought that growth of the state funded academic and research institutions will be better if they are taken over by the central government. Moreover, nurturing of such institutions is also the mandate of the center as per Indian constitution. Being geographically located in the Uttrakhand, they will actively participate in the growth of the newly formed state with financial support coming from the center.

In February 2001, the Uttaranchal government stating its inability to commit adequate funds for the functioning of the Observatory and wrote to the DST for its takeover. In order to study this proposal, the DST, in turn, formed a 10-member committee with Prof. V.S. Ramamurthy, then Secretary, DST as its Chairman and internationally reputed Scientists like Prof. J.V. Narlikar, then the Director, IUCAA, Pune; Prof. G. Srinivasan, then the President, Astronomical Society of India (ASI), Raman Research Institute (RRI), Bangalore and Prof. R. Cowsik, then Director, Indian Institute of Astrophysics (IIA), Bangalore as expert members. The committee unanimously recommended conversion of the state government Observatory into an independent autonomous research institution under the DST like RRI and IIA. Prof. M. M. Joshi approved this recommendation and the relevant parts from the approval note dated November 27, 2003, sent to the Union Cabinet for its consideration are reproduced below:

> … The Nainital Observatory was an initiative taken in 1954 by the Government of Uttar Pradesh for developing facilities in modern astrophysical research and carrying out frontline research in astronomy and astrophysics. The Observatory not only carried out several studies of stellar population and variable stars, but also functioned as recognized research center for training students of a large number of universities and educational institutions in this country. A Committee headed by Secretary, DST and consisting of internationally reputed Scientists like Prof. J.V. Narlikar, Prof. R. Cowsik and Prof. G. Srinivasan have strongly recommended for a major modernization of this Observatory with the addition of a larger telescope and also setting up modern monitoring facilities relating to global climate change. A suitable site has also been identified at Devasthal. But, unfortunately, the Government of Uttranchal has not been able to provide adequate support for well-known reasons for these programmes. Prof. Narlikar of the IUCAA had strongly advocated for granting autonomy to this Centre so that it is capable of growing to its full potential and this view has been accepted by the scientific community at large.

Considering that this is the only major research institution in the field of astronomy and astrophysics in the Northern part of the country and has also a strategic significance, I strongly believe that we should take all necessary steps to modernize this laboratory and integrate it in the national network of R&D institutions in related fields. The region also has several educational institutions like the Kumaon University, the Garhwal University, IIT Roorkee having a rich tradition of astronomy and astrophysics research and modernization of this laboratory will contribute substantially to upgrade the level of education in the region as a whole.

It is therefore, necessary to convert this Observatory into a National Level Research Institution under the Department of Science and Technology, Government of India and integrate it with the other institutions of the country in this area of research.

The Draft Cabinet Note (DCN) has already been circulated to the Government of Uttaranchal, Planning Commission and the other Ministries under the Government of India and their suggestions have also been incorporated in the revised DCN. The comments of the Department of Expenditure, Ministry of Finance and DST's clarifications on the same on pages 1 and 2 be added to it. Accordingly, Cabinet note is approved for being placed before the Cabinet along with my forgoing observations.

Based on this approval, the DST prepared final Cabinet Note which was approved in the Cabinet Committee meeting held on January 7, 2004. After the meeting, media was briefed about the state government proposal for converting the Observatory into an autonomous research institution under DST. In compliance of this Cabinet approval, ARIES was formed on March 22, 2004 and author became its founder Director and continued till his superannuation in the year 2014. During 2017 to 2018, Dr. Anil Kumar Pandey was Director of ARIES for over 1.5 years. Present Director of ARIES, Prof. Dipankar Banerjee, took charge on December 12, 2019. Fig. 3 shows photos of all three Directors of ARIES.

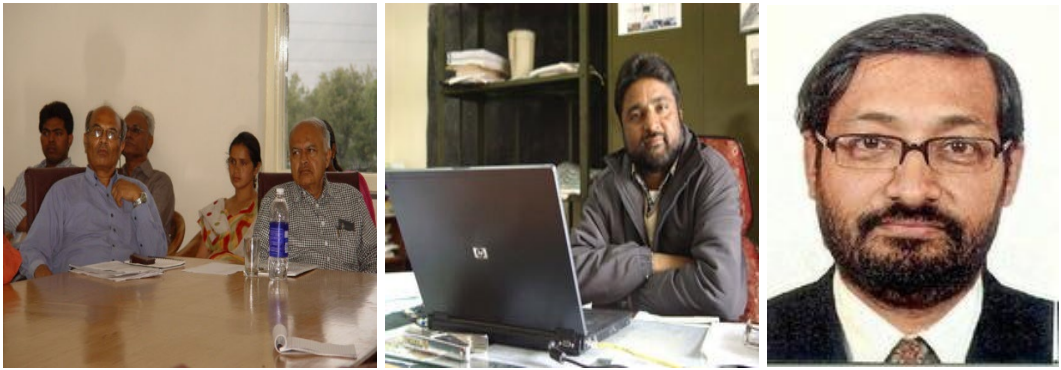

**Fig. 3** Left: Ram Sagar with Jayant V. Narlikar who delivered a talk at ARIES on June 26, 2009. A.K. Pandey (Center) and Dipankar Banerjee (Right) are also seen.

## 3 Evolution of Science and observational facilities during seven decades of ARIES

The research activities of ARIES can be broadly classified in three groups namely Galactic and extra-galactic, Sun and solar system and Earth's atmospheric sciences. The evolution of these researches is presented below along with information about associated observational facilities as both are inter-linked.

### 3.1  Galactic and extra-galactic research

In the field of Galactic and extra-galactic astronomy, research is based on observations of celestial bodies taken with the back-end instruments mounted on an optical telescope. For such observations, optical telescopes installed by the institute during its entire legacy of seven decades are listed in Table 1 (Sinvhal, 2006; Sanwal et al., 2018; Sagar, 2020 and references therein). The 79/51 cm Baker-Nunn (BN) camera was a special purpose Schmidt telescope (Bappu, 1958). Only 25-cm telescope is refractor type while other telescopes are reflector type. The optical elements of all telescopes are made of low-thermal expansion glass except in the case of International Liquid Mirror Telescope (ILMT) where primary mirror is made of liquid mercury (Surdej et al., 2018). The focal plane instruments are mounted at the prime focus of the ILMT and BN camera and at the Ritchey-Chretein (RC) Cassegrain or Coude or Nasmyth foci of other telescopes (Table 1). All telescopes, except the ILMT, can observe celestial objects located in any part of the sky. F-ratio is ratio of the telescope focal length (F) and its clear aperture. The plate scale is defined, in units of arc-sec/mm, as 206265/ F(mm). It connects the angular separation between celestial objects with their linear separation at the focal plane image of a telescope. Fast systems covering large part of the sky at the focal plane of a telescope have larger value of plate scale while vice-versa is the case for smaller value of plate scale.

**Table 1** The optical telescopes installed by the institute with aperture size and thumbnail image of the telescope, its manufacturer and year of installation are given in Column 1, 2 and 3 respectively.

| Aperture (cm) | Fabricator | Year of installation | Mounting | Focus | F-ratio | Plate scale (arcsec/mm) | Publications |
|---|---|---|---|---|---|---|---|
| 25 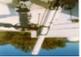 | Cook, UK | 1955 | Equatorial | RC | 15 | 55 | 14 |
| 38 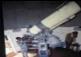 | Fecker, USA | 1960 | Equatorial | RC | 15 | 36 | 79 |
| 52 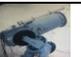 | CHT, UK | 1961 | Equatorial | Nasmyth, Coude | 13&70 | 31&6 | 41 |
| 58 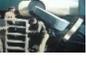 | CHT, UK | 1968 | Equatorial | Cassegrain | 15 | 24.6 | 17 |
| 79/51 BN 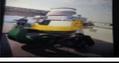 | SAO, USA | 1958 | Altitude-Azimuth | Prime | 1 | 401 | 2 |
| 104 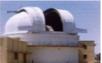 | Carl Zeiss, East Germany | 1972 | Equatorial | RC, Coude | 13&31 | 15.5&6.3 | >450 |
| 130 DFOT 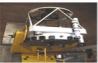 | DFM, USA | 2010 | Equatorial | RC | 4 | 40 | >100 |
| 360 DOT 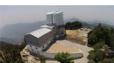 | AMOS, Belgium | 2016 | Altitude-Azimuth | RC | 9 | 6.4 | >70 |
| 400 ILMT 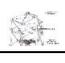 | AMOS, Belgium | 2021 | Zenith | Prime | 2 | 25.8 | >4 |

The 79/51 cm Baker-Nunn (BN) camera was a Schmidt telescope and used mainly for optical photography of the artificial Earth satellites. It was installed and operated in collaboration with Smithsonian Astrophysical Observatory (SAO), USA. Both 52-cm and 56-cm optical telescopes were supplied by the Cox, Hargreaves & Thompson (CHT), Ltd., United Kingdom (UK). Mounting, focal plane used for observations, F-ratio and plate scales of the telescopes are given in Column 4, 5, 6 and 7 respectively (Table 1). The RC in column 4 denotes the RC Cassegrain system. The number of publications related with the telescope are given in the last column. The 130-cm Devasthal fast optical telescope (DFOT), 360-cm Indo-Belgian Devasthal optical telescope (DOT) and 400-cm ILMT were installed at Devasthal peak while all others were installed at Manora peak, main campus of ARIES. The Advanced Mechanical and Optical System (AMOS), Belgium designed and built both 360-cm DOT and 400-cm ILMT. Till 1972, 6 optical telescopes of diameter ranging from 25 cm to 104 cm were installed at Manora peak while after a gap of ~ 4 decades, 3 modern telescopes were installed at Devasthal peak.

As depicted in the left panel of Fig. 4, the light-gathering powers of different-sized optical telescopes can be compared by calculating the ratio of their diameters squared. The advantage of collecting more light photons with a larger-aperture telescope is that one can observe fainter celestial objects. For sky background limited imaging observations (Sagar, 2000; Sagar, 2017), signal-to-noise (S/N) ratio at a frequency, ν, is

$$\alpha \ \sqrt{(A_{eff} \times I_t)/(\varepsilon D \times B(\nu))},$$

where $A_{eff}$ is the telescope light gathering power, including the losses due to optics and the quantum efficiency (QE) of the detector used at the focus of a telescope; $B(\nu)$ is the sky background intensity at frequency $\nu$; $I_t$ is the integration time and and $\varepsilon D$ (for telescopes of aperture $\geq 15$ cm) is the solid angle formed by Earth's atmospheric seeing. At a given astronomical observational site for a fixed $I_t$, deep imaging capacity of a telescope is therefore $\propto (A_{eff}/\varepsilon D)$.

This means that a 1-m telescope located at a $\varepsilon D = 0.5"$ site can provide as deep imaging of the celestial objects as a 2-m telescope located at a $\varepsilon D = 1"$ site, if other conditions are similar. Consequently, due to relatively dark sky and better atmospheric seeing, telescopes located at Devasthal peak provide deeper imaging of celestial bodies in comparison to similar telescopes located at Manora peak.

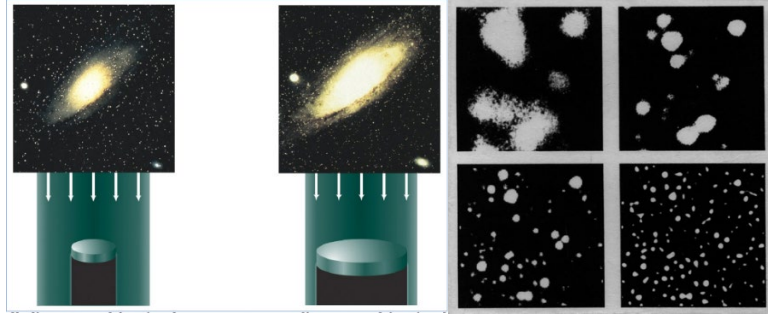

**Fig. 4** The light-gathering powers of different-sized optical telescopes (left). Four images of a central region of Omega Centauri (right panel).

Left panel in the Fig 4 shows two telescopes of different aperture sizes receiving photons from celestial bodies. More number of photons is received by the larger-sized telescope due to its larger surface area. The light-gathering powers of different-sized optical telescopes can therefore be compared by calculating the ratio of their diameters squared. The four images in the right panel, borrowed from Wilson (1989), are 12" x 12" field located near center of the southern galactic globular cluster Omega Centauri (NGC 5139). They show importance of improving angular resolution of an astronomical image. The first (upper left) picture was taken in 1984 with the 1-m European Southern Observatory (ESO) Schmidt telescope having plate scale of 67.5" per mm. During observations, seeing was ~2" and unsensitized, blue sensitive IIIa-J photographic emulsion was exposed for 10 min. Next upper right photo shows an excellent image obtained in 1977 at the Cassegrain focus of the ESO 3.6-m telescope having plate scale of 7.2" per mm. During observations, seeing was ~1" and photographic emulsion IIIa-J was exposed for 6 min 15 sec. The lower left picture is a 10 sec raw charge coupled device (CCD) image taken with the 3.58 ESO New Technology Telescope (NTT) on March 23, 1989 during NTT first light under excellent atmospheric seeing of ~0.3". The plate scale of the image is 5.4" per mm and the value of full width at half maxima (FWHM) estimated directly from stellar images is 0.33". This raw digital CCD image, after sharpening using advanced image processing techniques, is shown at the lower right. The FWHM value is now improved to 0.18"; the sharper stellar images are noticeable and fainter stars are detected with better S/N ratio.

Resolving power (RP) of ground based optical telescopes depends mainly upon the value of εD during the observations, plate scale of the telescope and spatial resolution of astronomical detector used at the focus of the telescope. In order to demonstrate importance of RP, 4 astronomical images of size 12" x 12" located in the central region of globular star cluster Omega Centauri (NGC 5139) are shown in the right panel of Fig. 4 (Wilson, 1989). Both upper images are photographic but seeing conditions during the observations are significantly different. Plate scales of the telescopes used for taking these images are also quite different. In the case of left lower CCD image, atmospheric seeing was excellent and exposure time was short. So, well resolved and faint stellar images can be clearly seen. An inter-comparison of the images indicates significant improvement in the RP i.e., angular resolution of CCD images. For example, one can see the triple star located in side red circle in all 4 images. Angular distance between the two closet components is ~0.79". The ESO Schmidt image does not indicate any multiplicity, the ESO 3.6-m barely resolve the system while the ESO 3.58-m NTT shows the three components well resolved from each other. Since the light is more concentrated on the CCD detector, the higher angular resolution also leads to detection of fainter stars. Better atmospheric seeing at Devasthal peak in comparison to Manora peak can, therefore, provide relatively deeper imaging of celestial bodies with the telescopes located at Devasthal.

For tracking celestial objects, 25-cm telescope used gravity drive (see detailed caption of Fig.5) while all other telescopes used stabilized electric power devices. Till the advent of CCD, photon detector on all telescopes was photomultiplier tube except on the 79/51 cm BN camera where highly sensitive emulsion photographic plates, available at that time, were used for taking photographs of Earth artificial satellites. After serving their purpose and also with installation of better and modern observing facilities, 25-cm, 38-cm, 52-cm, and 56-cm telescopes are now de-commissioned along with BN camera. However, due to their historical importance, they are described below in brief along with astrophysical studies carried out with the telescopes.

### 3.1.1 25-cm refractor and 38-cm, 52-cm and 56-cm reflector optical telescopes

Technical details of the institute's telescopes are given in Table 1 along with their year of commissioning. Till late 1980's, photons collected by the telescopes were detected by photomultiplier tubes, best astronomical detectors available at that time. The photometric observations taken with the telescope were recorded on strip chart using DC amplifiers and related electronic devices.

The 25-cm refractor was the first optical telescope installed initially at Devi Lodge and later on at Manora peak (1st and 2nd panel from left in Fig. 5). Both Visual and recorded observations of different type of celestial bodies started with the operation of this telescope. After installation of the telescope, a photoelectric photometer with 1P21 photomultiplier tube, designed and developed by Dr Vainu Bappu, was mounted on this telescope. It obtained the first photoelectric astronomical observations in Asia. Right panel of Fig. 5 displays photograph of chart recorded observations of an occultation of star BD -5° 5863 by minor planet asteroid Pallas on October 2, 1961. Duration of the occultation was estimated to be 25.53 second by Sinvhal et al. (1962). The photoelectric B and V photometric observations of different type of celestial objects including galactic clusters and variable stars were obtained. Variable stars' studies included β Canis Majoris stars, eclipsing variables and short period Cepheids. In some galactic clusters, a survey of H-γ absorption line intensities in early type stars was made. The telescope was also used for photoelectric measurements of the magnitude and polarization of the nuclei of comets Arend-Roland 1956h and Markos 1967d (Sinvhal, 2006).

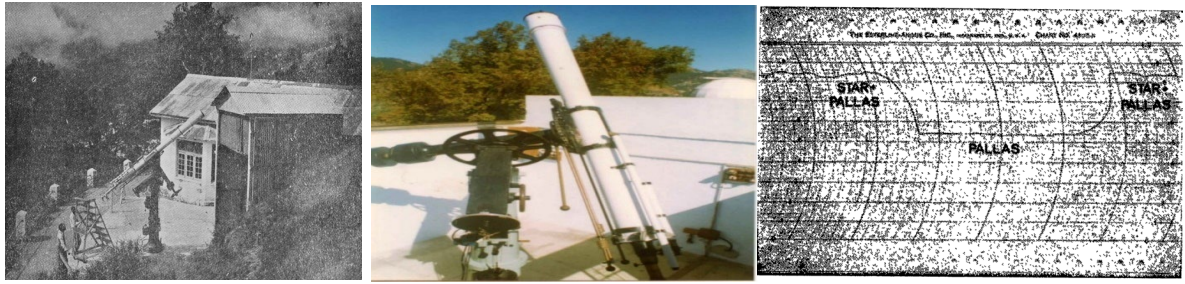

**Fig. 5 Left**: 25-cm Cook visual refractor telescope installed at Devi Lodge in 1955; **Center**: its photograph at Manora peak; **Right**: chart-recording during occultation of star BD -5° 5863 by Pallas on October 2, 1961.

Left photo in the Fig. 5 shows the 25-cm Cook visual refractor telescope installed at Devi Lodge in 1955. Central panel shows its photograph at Manora peak. Mechanical system of the telescope polar axis is mounted on a single pier equatorial German mounting. Main tube of the 25-cm f/15 telescope equipped with 2 finder telescopes (size ~10-cm and ~2.5-cm) is more than 4-m long. This and its counter weight, located on opposite side of the pier, are joined by a mechanical structure of the telescope Declination axis. In both photos, a weight tied with rope, mounted on a mechanical system below Declination axis, can be seen hanging towards lower part of the pier. The falling of this weight due to gravity drives Right Ascension axis of the telescope for tracking the celestial objects. Hence, it is called gravity driven telescope. Right panel shows photograph of chart-recording during occultation of star BD -5° 5863 by Pallas on October 2, 1961.

Column 1 of Table 1 shows thumbnail photographs of the 38-cm, 52-cm and 56-cm reflectors successfully installed at Manora peak, Nainital. During 1961 to 1997, the 38-cm reflector telescope was used for the observations of δ Scuti stars, β Canis Majoris stars, and eclipsing binaries. From 1998 to 2001, the telescope was shifted to the Devasthal site for seeing measurements using modern differential image motion monitor (DIMM) technique (Sagar et al., 2000). The main instrument on the 52-cm reflector telescope was a low-resolution laboratory spectrophotometer. The energy distribution studies of Cepheids, Be stars and comets were carried out during 1964 to 1995. In December 1996, the telescope was shifted to the Devasthal site for carrying out seeing measurements using DIMM and atmospheric extinction measurements. After successful completion of the project, the telescope was decommissioned in the year 2002. The 56-cm reflector telescope was used for photoelectric photometry and spectrophotometry observations of celestial objects. After a few years, the primary mirror of the telescope developed blemishes. It was, therefore, replaced by a new primary mirror made in the institute optics shop. Photometry of flare stars and spectrophotometry of comets were done with this telescope till formation of ARIES. As listed in Table 1, maximum numbers of research papers were published with the observations taken with the 38-cm reflector telescope.

### 3.1.2 Optical tracking of artificial earth Satellite

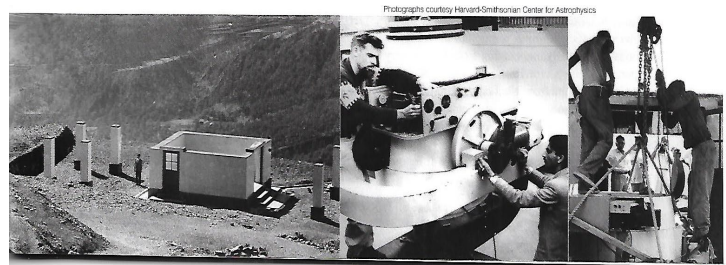

**Fig. 6** BN camera station under construction at Manora peak, Nainital, 1958.

Left panel in Fig. 6 shows the BN camera station under construction at Manora peak, Nainital in 1958. In the central photo, young Dr. C.D. Kandpal (right), UPSO and Mr. Samuel Whidden, SAO, are making final adjustments on the BN camera while in the right photo, BN camera being unloaded at the Manora peak, Nainital. Dr. M.K. Vainu Bappu is seen with a camera next to the fork while at the doorway to the left is Dr. S.D. Sinvhal. Photo Courtesy: Satapathy (2005).

In collaboration with the SAO, USA, the BN camera was installed at Manora peak, Nainital, main campus of ARIES. Its aim was to photograph the artificial Earth satellites against the starry sky background. By measuring relative coordinates of the satellite image with respect to the surrounding star images, Right Ascension and Declination of the satellite can be determined for the tracking time. This yields the direction from the camera station to the satellite position (orbit point). The measures of position of the satellite at various known instants are the basic data needed for deriving its orbital characteristics. Various perturbation effects like atmospheric drag, gravitation anomalies and the oblateness of the Earth affect orbital parameters of the satellite. It is, therefore, possible to study the nature of these perturbations and the secular changes in the orbital characteristics with long term precise observations of the satellite's positions.

The BN camera of a 79-cm aperture primary and a 51-cm diameter corrector disc provided a focused 30°× 5° curved field at the prime focus. Installation of this camera along with a precision timing system capable of recording up to ten millionth part of a second was completed by August 1958. This was part of a global network of 12 such BN camera stations set up by the SAO.

Others were located in New Mexico, Florida, Hawaii, Japan, Iran, Spain, South Africa, Australia, Curacao, Argentina and Peru.

The first track of an artificial satellite recorded by this BN camera was in the year 1958. Due to good sky conditions and dedication of the staff, quite often, the success percentage of tracked artificial Earth satellite transits at the Manora peak was the highest among the twelve global SAO stations. At the end of the collaboration in 1976, SAO gave the BN camera observing facility to the institute. Recently, it's optics have been used in building a Schmidt telescope with a CCD prime focus imaging device. Over 45,700 satellite transits, including those of India's Aryabhatta and Apollo-11, 12 and 17 of the USA, were successfully recorded from Manora peak (Satapathy, 2005; Sinvhal, 2006). Out of them, two sample photographs, are shown in the 1st and 2nd panel of Fig. 7. As a result of this activity, the position of the BN Camera is determined to an accuracy of better than 10 meters on a world datum (Longitude=79°27'25.5" East, Latitude=29°21'39" North, Altitude=1927 meter). It thus became amongst the few locations on earth with coordinates so precisely determined. Around 1975, the Survey of India therefore tied their triangulation network to this benchmark.

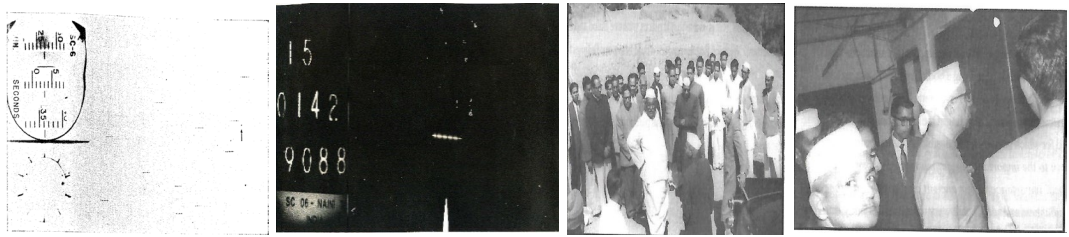

**Fig. 7** Left: photograph of an Earth artificial satellite taken on October 4, 1958 with the BN camera; Center and Right: Pt. Nehru, Sampurnanand and Shastri with several other dignitaries at BN camera station in 1959.

Left panel in Fig. 7 shows a photograph of an Earth artificial satellite taken on October 4, 1958 with the BN camera (Bappu, 1958). The time presentation is seen on the left of the picture and reads 23 hrs 23 min 3·3612 sec. The oscilloscope yields fractions of 10 msec. The angular velocity of the satellite was 541" per second of time. The total exposure on the satellite equals 1.6 sec. The satellite image is marked by an arrow. The photo displayed in the 2$^{nd}$ panel (from left) is trailed image of a satellite taken against dotted images of background stars at different epoch. In this case, BN camera tracked the stars and time was printed on the image. Two photos shown in 3rd and 4th panel (from left) were taken on June 18, 1959 during visit of former and first Prime Minister of India, Pt. Jawaharlal Nehru, Ex-Chief Minister of UP, Dr. Sampurnanand, Mr. Lal Bahadur Shastri and several other dignitaries to the BN camera station, located at Manora Peak, main campus of ARIES. The team led by Dr. Bappu showed this newly built international observing facility to the dignitaries.

### 3.1.3 Fifty golden years of the 104-cm Sampurnanand telescope

The dome of 104-cm telescope was designed, fabricated and installed by M/s Triveni Structural Naini, Allahabad (now Prayagraj). In this dome, the 104-cm telescope, acquired from Veb Carl Zeiss Jena, was installed in February 1972 (photo in left panel of Fig. 8) and observations with the telescope started in October 1972. Details of the telescope optical, electrical, mechanical components, history of procurement and installation at highest point of the main campus of ARIES are very well documented by Sinvhal et al. (1972). Briefly, the telescope has equatorial English mounting and rests on North and South piers. It is housed in the first floor of a two-story building. Three finder telescopes (26.4 cm, f/14, reflector; 20 cm, f/15, refractor and 11 cm, f/7, refractor) are also mounted on the 104-cm telescope's main tube. The telescope was formally dedicated to the astronomical community on June 7, 1973 by Prof. M.G.K. Menon, then Secretary, Department of Space, Government of India. His Excellency Sri Akhbar Ali Khan, the then Governor, UP presided over the function. Other dignitary present in the function was Sri Swami Saran, Minister for Civil Aviation & Scientific Research, UP government. To honour germinator of the institute, the telescope was named after Dr. Sampurnanand. The images in the 3rd and 4th panel (from left) of Fig. 8 show pictures of galactic globular cluster M13 and face-on spiral galaxy M51 respectively. They were taken with the 104-cm telescope using 2Kx2K CCD camera having 13'x13' field of view (FOV). During last 5 decades, this telescope has extensively been used for optical and near infra-red (NIR) photometric, spectroscopic and polarimetric observations of celestial objects and events using back-end instruments, described in details by Sagar et al. (2014), are briefed below.

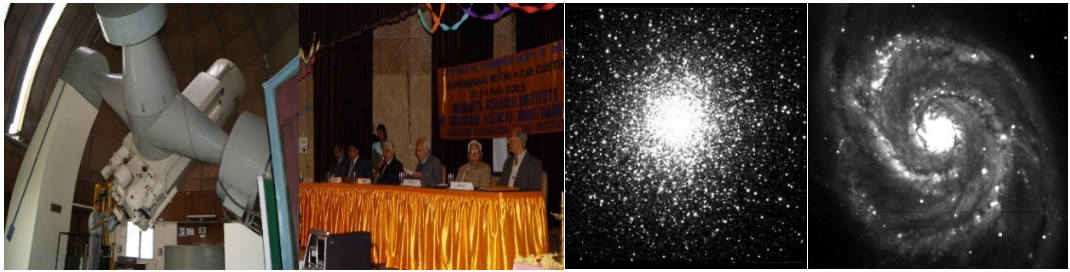

**Fig. 8** Left: the104-cm Sampurnanand telescope, Cetnter: Delegates at one day seminar at ARIES in 2005; Right: winding arms of the M51, located at a distance of ~31 million light-years from Earth.

Photo of the104-cm Sampurnanand telescope is shown in the left panel of Fig. 8. Right Ascension (Polar) axis of the telescope is mounted on the North and South piers. This axis points towards north celestial pole (close to North pole star). Main tube of the telescope and its counter weight are mounted on the declination axis which is perpendicular to the polar axis of the telescope. Three finder telescopes mounted on the main telescope tube can be seen in the picture. Photo displayed in the 2$^{nd}$ panel (from left) was taken during inaugural function of a one-day international meeting organized by ARIES on star clusters along with the 23rd meeting of the ASI held during 21 to 24 February 2005. From left to right, Dr. K. Sinha, Dr. Ashok Ambastha, Prof. S.N. Tandon, Prof. D. Sharma, Dr. S.D. Sinvhal and author are sitting on the dais. CCD images of the Galactic globular cluster M13 and the Whirlpool galaxy M51 taken with the telescope are displayed in 3rd and 4th panel (from left) respectively. The M13 (~12 Gyr old) is located in the constellation of Hercules at a distance of 22.2 K light-years from us. Due to its large distance, stars in the central region of the cluster are not resolved. In the right most picture, the winding arms of the M51, located at a distance of ~31 million light-years from us in the constellation Canes Venatici, appear like a grand spiral staircase sweeping through space. They are star-formation factories, compressing hydrogen gas and creating clusters of new stars. Face-on view of the galaxy allows astronomers to study its structure and star-forming processes.

Back-end instruments, used till late 1980's, were a Cassegrain plate holder, Meinel camera, UBV photoelectric photometer, NIR photometer, low resolution laboratory spectrophotometer and Cassegrain spectrograph. NIR photometer built at ARIES employs InSb based photovoltaic detector and a pre-amplifier electronics. In the year 1989, with the efforts led by Dr. Vijay Mohan, a 384x576 pixels$^2$ CCD camera system was acquired and installed for the first-time as a new back-end instrument (Mohan et al., 1991). QE of a CCD is very high in comparison to photomultiplier and photographic plates used earlier and also, being a 2-dimensional detector, CCD acts like a N star photometer. Use of CCD, therefore, made a quantum jump of ~ 200-fold in the observing capabilities of the 104-cm telescope. In the year 1991, SBIG ST-4, a small CCD camera system, was installed on the 20-cm finder telescope for auto-guiding the telescope during long exposures. For spectrophotometric observations, an optical multichannel analyser with 1-dimensional 1024-pixel reticon array as detector was also acquired and installed in the year 1991. Large CCD camera systems having 1024x1024 and 2048x2048 pixels$^2$ were installed in the year 1993 and 1998, respectively. A 3-channel fast photometer became operational in the year 2000 and a polarimeter named ARIES Imaging POLarimeter (AIMPOL), developed in-house by Rautela et al. (2004), became operational in the year 2004. Galactic open clusters, binary stars, Be stars, quasars/blazars, occultations by planets and asteroids, optical afterglows of gamma-ray bursts, Novae and Supernovae, active galactic nuclei, rapidly oscillating Ap (ROAp) stars, IRAS sources, stellar interferometry, microlensing events towards M31 were the main research topics of interest. A large number of scientists from India and abroad have used the telescope. The observations taken with this telescope have contributed to a total of over 450 research publications and about 40 PhD Theses so far. All these clearly demonstrate extensive use of the 104-cm Sampurnanand telescope during last 5 decades. A workshop on *Astronomy with moderate size optical telescopes* was organized during April 7-9, 1997, to commemorate the silver jubilee year of the 104-cm Sampurnanand telescope. Shri Prabhat Chandra Chaturvedi (IAS) Secretary, DST, UP and His Excellency Mr. Romesh Bhandari, the then Governor, Government of UP inaugurated the event. The workshop was attended by around 120 participants from all over the country and a total of 65 presentations covering wide range of topics were made. Its proceedings were published in *Bulletin of ASI*, 26 (1998): 351–614.

The Golden Jubilee year celebration of the Institute's founding was celebrated from May 2004 to April 2005. As a part of it, ARIES hosted one day international meeting on **Star Clusters** along with the 23rd meeting of the ASI from 21 to 24 February 2005. Second panel (from left) of Fig. 8 shows a snapshot photo taken during the inaugural function of the event. Star Clusters is one of the topics to which the ARIES has made major contributions over many years. The meeting attracted a large audience from all over India, together with a number of overseas experts in star cluster research. A summary of the scientific work presented in the meeting is given by Cannon (2006) while some talks are published in *Bulletin of ASI*, 2006, Vol 34.

### 3.1.4 Devasthal Observatory hosting 1.3-m DFOT, 3.6-m DOT and 4-m ILMT

After an extensive site characterization carried out using modern instruments over an extended period (Sagar et al., 2000; Stalin et al., 2001), ARIES identified Devasthal as a potential site for locating modern optical telescopes. In comparison to Manora peak, atmospheric seeing at Devasthal is better at least by a factor of 2. A comparison of imaging observations obtained with optical telescopes installed at Devasthal and Manora peak indicates that former detects celestial objects at least 4 times fainter in comparison to later (Sagar et al., 2019, 2020). This is (as discussed above in Section 3.1 and Fig. 4) mainly due to better atmospheric seeing and dark sky conditions at Devasthal in comparison to Manora peak. A brief historical description of modern telescopes installed at Devasthal is given below.

Prof. K. Kasturirangan, then member Planning Commission and Chairman, Governing Council, ARIES laid foundation stone of the 1.3-m DFOT building on April 9, 2006. Devasthal site became an observatory in October 2010 when this telescope was successfully installed and started observing celestial objects. The control system of the telescope is capable of operating the telescope automatically and can also be interfaced with the standard sky-viewing software such as The Sky, eliminating the need for any finding chart. Dr. T. Ramasami, then, the Secretary, DST, dedicated the telescope to the nation on December 19, 2010. Technical details of the telescope and backend instruments and their performance verification results are published by Sagar et al. (2011). The telescope is capable of imaging wide circular area (~ 66' diameter) of sky justify its name DFOT acronym for Devasthal fast optical telescope. First light CCD images of Orion star-forming region, open star cluster M67 of our galaxy, the galaxy NGC 598 and M82 are shown in the right most panel of Fig. 9.

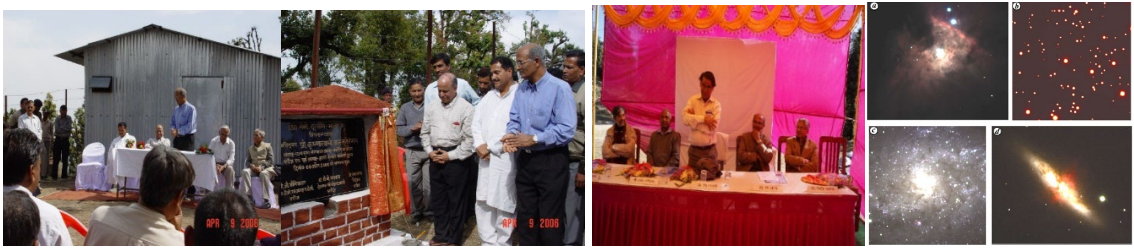

**Fig. 9** Foundation stone laying ceremony of 1.3-m DFOT by Prof. K. Kasturiranagan, Chairman, Governing Council, ARIES, on April 9, 2006.

Fig. 9 both (left and next right) photos show snapshot of foundation stone laying ceremony of 1.3-m DFOT by Prof. K. Kasturiranagan, then Chairman, Governing Council, ARIES, on April 9, 2006. From right to left, Mr. Ashish Ganju, Prof. G. Srinivasan, author (addressing the gathering), Prof. Kasturirangan and Mr. Sher Singh Naulia, Block head of the Devasthal region, are seen in the first left photograph. In the next right photo, curtain raising of the foundation stone can be clearly seen. Third photo from left shows a snapshot of dedication ceremony of the telescope to the astronomical community on December 19, 2010 by Dr. T. Ramasami, then, Secretary, DST. On the dais, from left to right, Prof. Dipankar Bhattacharya, Shri A.B. Koutilya, Dr. T. Ramasami (addressing the gathering), author and Prof. P.C. Agrawal are present. In the right most panel, 4 first light images taken with the 1.3-m DOFT using 2Kx2K CCD camera are displayed (Sagar et al., 2011). Upper left marked (a) is a broad-band BVR colour composite image of the famous Orion star-forming region, also known as 'mrigshirsha nakshatra'. Upper right marked (b) is an image of M67 open star cluster in white light, while lower left marked (c) image shows the galaxy NGC 598 in BVR broad-band colour. Lower right marked (d) image shows ionized gas seen in H-alpha (red colour) in a starburst galaxy M82.

In March 2007, ARIES awarded the contract to the AMOS for design, manufacture, integration, testing, supply and installation of a 3.6 m aperture size alt-azimuth technologically advanced modern thin mirror optical telescope at Devasthal (Kumar et al., 2018; Sagar et al., 2019). Integration and first-light verification tests of the telescope were carried out at the AMOS factory in the year 2012. Weight of the whole telescope is ~150 MT while maximum weight of a single part of the telescope is ~14 MT. Whole parts of the telescope were transported from the AMOS factory in Belgium to the Devasthal site, Nainital in India during the year 2013. Kumar et al. (2018) have provided details of construction of technically complex telescope dome enclosure as well as design and fabrication of overhead cranes for integration of the entire telescope. The 3.6-m DOT was installed successfully during October 2014 to March 2015. Engineering verification and rigorous on-sky tests of the telescope were performed during May 2015 to February 2016 using an air-cooled Micro-line ML 402ME CCD test-camera mounted at the axial RC Cassegrain port of the 3.6-m DOT (see Omar et al., 2017; Kumar et al., 2018; Sagar et al., 2019 for details). The chip size of the test-camera was $768 \times 512$ pixels$^2$ while its plate scale at the telescope focus was $0''.06$/pixel. A few sample images of celestial objects taken during telescope acceptance tests are displayed in the 3rd, 4th and 5th panel (from left) of Fig. 10. The TIRCAM2 camera (Ojha et al. 2018; Baug et al. 2018 and references therein) was used to characterize NIR sky properties of the Devasthal for the first time. Fig. 11 shows K-band image of the Sh 2 -161 region taken on May 23, 2017 with this camera.

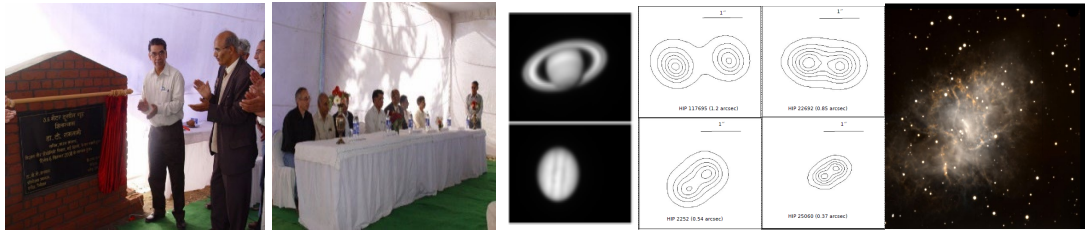

**Fig. 10** The foundation laying ceremony of the 3.6-m DOT building by Dr. T. Ramasami, Secretary, DST.

First two photos from left show snapshot picture of foundation laying ceremony of the 3.6-m DOT building by Dr. T. Ramasami, Secretary, DST, on September 6, 2008. During the ceremony he remarked that "*Devasthal will house one of the Asia's most powerful telescopes. We have made a small beginning for a giant step in observational science today. Let ARIES bring glory to India*". In the 2$^{nd}$ photograph (from left), Dr. Tom Richler (Germany), Prof. S. Ananathkrishnan, Prof. P. C. Agrawal, Dr. T. Ramasami, author, Mr. Somesh C. Jhingan and Prof. T. P. Prabhu are sitting on the dais while Dr. B.B. Sanwal, then Project manager of the 3.6-m DOT, is addressing the gathering. Third (from left) photo display white light images of the planets Saturn and Jupiter taken with the test-camera on March 21, 2015 during engineering verification of the 3.6-m DOT. Fourth panel (from left) shows the iso-intensity contour images of four close binary stars, having angular separation between 0″.37 to 1″.2, observed during November–December 2015 with the test-camera mounted on the telescope. Binary stars having separation of sub-arc-second were clearly resolved in this figure. In one of the best observations, a binary star with known angular separation of ~0″.4 was well resolved on the night of November 30, 2015. Right most picture of angular size 6."5 x 6."5 is a 180 s exposed V-band image of the Crab nebula (a supernova remnant). It was taken on December 11, 2015 with the 4Kx4K CCD camera (Pandey et al., 2018) mounted at the main axial port of the 3.6-m DOT. Sharp and circular images indicate that the telescope is capable of providing good quality celestial images (Sagar et al., 2022 and references therein).

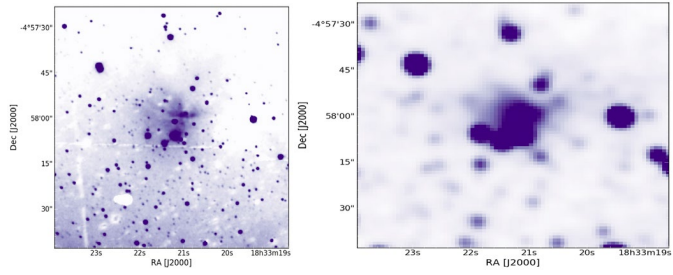

**Fig. 11** Deep K-band (2.19 µm) imaging of the Sharpless (Sh 2 -161) region (α J2000: $8^h33^m21^s$; δJ2000 = -04°58'02").

Deep K-band (2.19 µm) imaging of the Sharpless (Sh 2 -161) region (α J2000: $8^h33^m21^s$; δJ2000 = -04°58'02") was carried out on May 23, 2017 using TIRCAM2 (Ojha et al. 2018; Baug et al. 2018 and references therein) mounted at the Cassegrain main port of the 3.6-m DOT. The FWHM of the stellar images were ~0."5 (Sagar et al., 2022). The left panel is 80" x 80" TIRCAM2 K-band image, while the same field taken from 2MASS survey is shown in the right panel. A comparison of both images clearly indicates that TIRCAM2 image can resolve far more stars than 2MASS and observe dust and gas emission near the central region of Sh 2-61.

Being an Indo-Belgian venture, the 3.6-m DOT was technically activated jointly by the Premiers of both India and Belgium on March 30, 2016 from Brussels, Belgium. Since then, the telescope is in regular use by the astronomers from all over the globe. Both optical and NIR observations reveal that the performance of the telescope is excellent and deep images of celestial bodies with sub-arc-sec resolutions can be obtained with it (Sagar et al., 2022 and references therein). All these clearly indicate that the night sky at Devasthal is dark and qualities of imaging and angular resolution observations are comparable to other similar telescopes available world-wide.

Surdej et al. (2018) have given technical details and scientific objectives of the 4-m ILMT project. In this telescope, the surface of mercury liquid placed inside a cylindrical container that is in rotation around its vertical axis acts as a primary mirror since it takes the shape of a paraboloid under the constant pull of gravity and of the centrifugal acceleration. This thin paraboloid rotating layer of mercury focuses the light from a celestial object at its focal point. Consequently, such a telescope always observes at the zenith where observing conditions are the best from ground but can image only a small strip (~ 30') of zenith sky. The telescope was successfully installed at Devasthal recently and have started routine sky observations.

### 3.1.5 Historical account of main scientific results in the field of Galactic and extra-galactic studies

The observations taken with the telescopes of the institute have contributed to a number of historically and scientifically important results (see Sinha, 2005; Sagar, 2006 & Sagar et al., 2014 for details).

In the year 1955, the first photoelectric observations of celestial bodies from Indian soil were obtained at the institute. This observational technique was successfully used for observing India's fist occultation of a star by a minor planet Pallas on October 2, 1961. The 104-cm Sampurnanand telescope was used to carry out photoelectric observations of the Uranus occulting the star SAO 158687; BD-19°4222 and Hyd-20°51695 on March 10, 1977; April 26, 1981 and May 1, 1982 respectively. These observations contributed significantly to the discovery of rings around the Uranus. The photoelectric photometric observations of occultation of the star MKE 31 by Neptune on September 12, 1983 and of Saturn occulting the star SAO 158913 on 24 and 25 March 1984 and the star SAO 158763 on May 12, 1984 were taken with the observational facilities of the institute. These optical observations contributed to the discovery of rings around Neptune (Pandey et al., 1984) and additional rings around Saturn (Mahra et al., 1985 and references therein). Recently, a stellar occultation by Pluto was observed on June 6, 2020 with the 1.3-m DFOT and 3.6-m DOT telescopes, using imaging systems in the I and H bands, respectively. These precise observations allowed to put important constrain on the evolution of Pluto's atmosphere (Sicardy et al., 2021).

Based on optical tracking observations of Earth's artificial satellites taken during late 1950's with the BN camera as part of a global network of 12 stations, the location of BN camera at Manora peak has been determined with an accuracy of 10 m in the frame of reference of Standard Earth. This information is valuable for geological survey of India. In the year 1956, studies of eclipsing binaries started in the institute. About 50 eclipsing binaries have been observed and studied so far. Orbital elements, masses and radii of these binaries have been calculated. Such determinations are very useful for number of astrophysical studies.

From India, first polarisation and photometric measurements of two comets named Arend-Roland (1956h) and Markos (1957d) were carried out from the institute in the year 1957. However, after more than a decade, systematic spectrophotometric studies of comets began with the observations of Comet Bennett in 1970. From the observations of about 30 comets taken from the institute, more than 60 research papers have resulted so far. Spectrophotometric observations were used to study variation of production rates of observed species with heliocentric distance.

Using high angular resolution lunar occultation technique, both 1.3-m DFOT and 3.6-m DOT telescopes have been used to estimate angular diameter of cool and binary stars since March 2016 (Ref. Richichi et al., 2020 and references therein). These observations of milliarcsecond accuracy have provided, for the first-time, angular diameter of many cool stars and also resolved close binaries. Such measurements provide valuable insight in the formation and evolution of these stars. During 1995 to 2004, the precise photoelectric observations taken from the institute led to the discovery of four δ-Scuti and one ROAp star. During last decade, a systematic search for variable stars in star clusters has been carried out using existing facilities of ARIES. These observations discovered several new variables. They clearly demonstrate astrophysical importance of different types of variables present in a star cluster.

About 25 new northern hemisphere soft X-ray sources discovered in surveys with the Einstein and ROSAT satellites are being monitored during last 2 decades with ARIES telescopes using CCD cameras as N-star photometer. Optical variability has been established in five of them including discovery of two chromospherically active stars namely FR Cnc and HD 81032. Gamma Ray Bursts (GRB), originating far out in space, are short duration (<1 second to several minutes) most powerful events known in the Universe after the Big Bang. The first Indian optical observations of an afterglow of a GRB 990123 were made at ARIES on January 23, 1999. Since then, more than 30 afterglows have been successfully observed with the telescopes of ARIES. Amongst these, the earliest optical observations of GRB 000301C in R band and GRB 030329/SN 2003dh in UBVI bands have been carried out from ARIES. Recent very deep photometric observations taken with the 3.6-m DOT have contributed significantly. These observations in combination with the published one including those at other wavelengths are used to study the properties of GRB afterglows. The optical afterglow light-curves, spectral energy distributions and calculated energetics of these bursts are used for putting observational constraints on the popular GRB progenitor's models. Our observations support core collapse model for progenitor of long-duration GRBs. A good number of supernova and novae events have also been monitored with the observational facilities of ARIES including first photoelectric photometric observations of Nova Sagittarii taken in the year 1977.

Star clusters are ideal tool for studies related to star formation and stellar Evolution. Since 1970's, several galactic star clusters and a few globular clusters located in our Milky-way galaxy have been observed at optical and NIR wavelengths. Polarimetric studies of a number of star clusters have been carried out using observational facilities of ARIES. These observations combined with archival data have led to the studies of open clusters pertaining to their star formation efficiency, age distribution, mass function (MF) and luminosity function. Spatial structure and the interstellar extinction in young open star clusters have also been studied. The slope of the MF above 1 solar mass for young star clusters (age < 100 Myr) is found to be universal with a slope of Salpeter value within uncertainties of its observational determination. This program has given many well recognized results of fundamental importance in the field of star-clusters.

Optical Variability of Powerful Active Galactic Nuclei have been studied at ARIES since late 1990's. The sensitive observations to detect intra-night optical variability of milli-magnitude amplitudes in radio-quite Quasars are being carried out at ARIES. The observations obtained so far, indicate that at least some radio-quiet quasars do exhibit micro-variability, albeit somewhat less often, typically for shorter times, and usually less violently than that exhibited by radio-loud quasars and blazars.

In order to study dark matter in the galaxies, in 1998, ARIES started looking for possible gravitational micro-lensing events towards M31 galaxy under the Nainital micro-lensing survey project. The lensing is believed to be caused by massive objects present in the halo of the galaxies, which are considered as possible dark matter candidates. To search for micro-lensing events towards M 31, a technique called the Pixel Method has been developed in collaboration with a French group called AGAPE (Andromeda Gravitational Amplification Pixel Experiment). Four years of long optical observations taken from the institute have led to detection of one possible micro-lensing event.

### 3.2 Sun and Solar Physics

A 25-cm 6.5° off-axis f/66 skew Cassegrain telescope with the double pass horizontal grating spectrograph and the 46 cm Coelostat (first left panel of Fig. 12) were the basic instrument for solar observations in 1970's. It was decommissioned after spectroscopic observations of the Sun for about a decade. In order to record solar flares and prominences, a 25-cm coelostat and a 15-cm f/15 objective lens equipped with a Bernard Halle 0.5°A narrow pass band filter was used with a sensitive photographic detector and fast time recording system available at that time. In the year 1993, this observing facility was upgraded with modern fast time recorder and the CCD Camera systems mounted on a Carl Zeiss Jena 15-cm f/15 Coude tower telescope (2nd panel from left in Fig. 12). More details on these observing facilities are given by Gaur et al. (1998) and Sinha (2005). The solar tower telescope is equipped with fast CCD cameras and number of narrow band filters for imaging the Sun at different wavelengths. An image of solar flare 4B/Xl7.2 observed on October 28, 2003 with 15 cm solar tower telescope is shown in 3rd panel from left in Fig. 12. This was one of the largest flares of the solar cycle 23, which occurred near the Sun's center and produced extremely energetic emission almost at all wavelengths from γ-rays to radio-waves. In order to study physical processes responsible for this flare, these observations at optical wavelengths were combined with its space observations taken by SOHO, RHESSI and TRACE (Uddin et al., 2006). On June 8, 2004, transit of Venus across solar disk, a rare celestial event, was observed from ARIES. One of the snapshot images taken during transit is shown in the right most panel of Fig. 12.

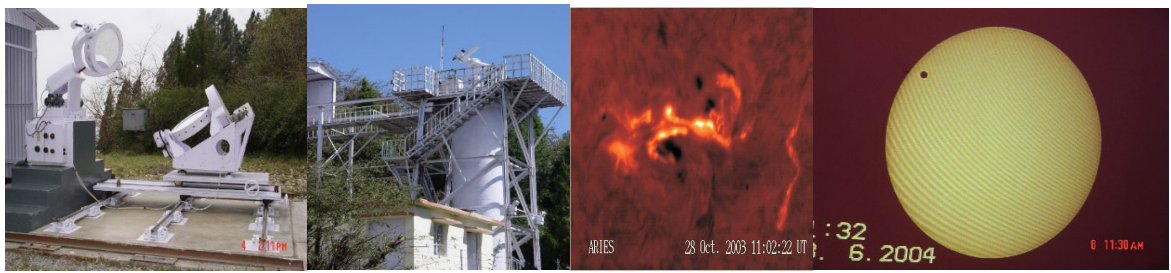

**Fig. 12** The 46-cm coelostat with the 15-cm f/15 Coude refractor telescope.

The 46-cm coelostat is shown in the 1st left panel while the 15-cm f/15 Coude refractor telescope (2nd panel from left) is mounted atop of a 9-m high tower for observations of solar activities. In the left panel, two mirrors of the coelostat system are seen in the photo. A flat mirror looking up on the right side of the photo is turned slowly by a motor to reflect the Sun light continuously into a fixed telescope. The mirror is mounted to rotate about an axis through its front surface that points to a celestial pole and is driven at the rate of one revolution in 48 hours. The telescope image is then stationary and nonrotating and can be used for its observations. The 15-cm refractor telescope is mounted on top of a tower so that effects of atmospheric turbulence arising from ground can be minimized while taking images of the Sun. Photo in the 3$^{rd}$ panel from left displays the image of the 4B/X17.2 class solar flare in Hα filter taken on October 28, 2003 with the tower telescope (Uddin et al., 2006). The right most panel of the figure shows a snap shot image of Venus (black dot) transiting the Sun taken on June 8, 2004 using observing facilities of ARIES.

In the early 1960's, studies related to solar molecules started at the institute. While multi-wavelength observational studies of transient phenomenon occurring on the Sun started in early 1990's after installation of fast recording systems on the solar telescopes. Sinha (2005) and Sagar et al. (2014) have given detailed account of these studies. The significant findings in the field of the solar research are: (i) using the Sun as a laboratory source, oscillator strengths have been determined for the MgH, $C_2$ and CN bands. Out of many new molecular species predicted by the solar group of the institute, the presence of SH, SiO and additional Swan band lines of $C_2$ have been confirmed observationally. (ii) Multi-wavelength observations of solar activities were used for modelling of impulsive flares and associated transient phenomena, statistical study of North-South asymmetry of the soft X-ray index, distribution of H-alpha areas, space-weather, etc. during solar cycles 21, 22 and 23. (iii) Multiple acoustic and sausage oscillations in the solar loops and the leakage of fast magneto-acoustic oscillations through the magnetic network core in higher corona were discovered. These will shed new lights on the dynamical study of solar atmosphere.

### 3.2.1 Expeditions to total solar eclipses

In order to observe total solar eclipse (TSE) of June 20, 1955, institute's first expedition went to Ceylon (now Sri Lanka) from Varansi under the leadership of Dr. M.K. Vainu Bappu. Other team members were Dr. S.D. Sinvhal and Sri Murari Lal Sharma, Lecturer at the Government Sanskrit College, Varansi. In order to study various parts of the corona by photographic polarimetry, a 10-cm f/38 lens and an 8-cm telescope were installed at the observation site in Hingurakgoda. A 15-cm f/4 camera on a tripod was also commissioned. Unfortunately, planned observations could not be carried out due to cloudy sky (Sinvhal, 2006). Similar was the case for another TSE expedition of August 11, 1999. Observational set up for this TSE observation is shown in left panel of Fig. 13. Institute successfully participated in TSE of February 16, 1980 from Palem, Mehboob Nagar, AP, India; November 24, 1995 from Meja Khas, Allahabad (now Prayagraj), UP, India; November 23/24, 2005 from Antarctica; March 29, 2006 from Manavgt, Turkey and July 22, 2009 from Anji, China. The TSE images taken by ARIES team during last two expedition are shown in central and right panel of Fig. 13. The high cadence imaging of solar corona during TSE provides the evidence of the role of fast magnetohydrodynamic waves in the heating of this mega-Kelvin solar atmosphere and help in solving the mystery of coronal heating.

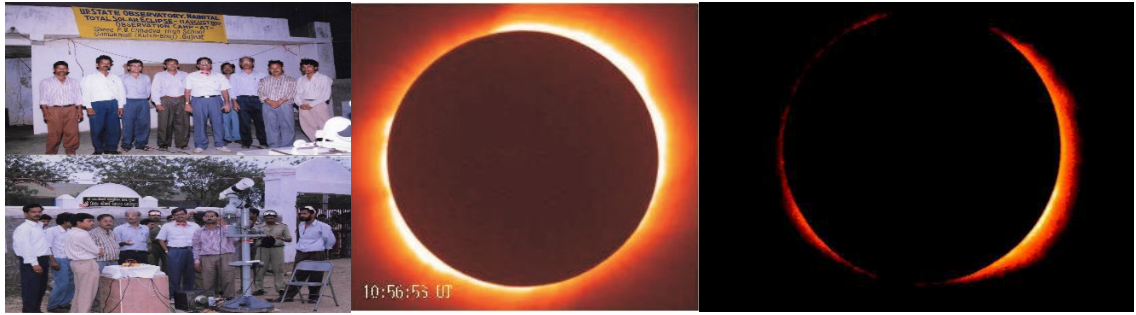

**Fig. 13** The observational set up installed in Kutch, Gujarat for observing TSE of August 11, 1999.

The left panel shows observational set up installed in Kutch, Gujarat for observing TSE of August 11, 1999, while photos in central and right panels show TSE images taken by ARIES team on March 29, 2006 and July 22, 2009 respectively.

During the year long (May 2004 to April 2005) Golden Jubilee celebration of the Institute's founding, ARIES hosted an International Solar Workshop on *Transient Phenomena on the Sun and Interplanetary Medium* during April 5-7, 2005. Its proceedings were published in the *Journal of Astrophysics and Astronomy* Vol 27 (2006) 57–372. The highlight of inaugural function was the release of "*Fundamentals of Solar Astronomy*" book authored by Prof. Arvind Bhatnagar and Prof. William Livingston.

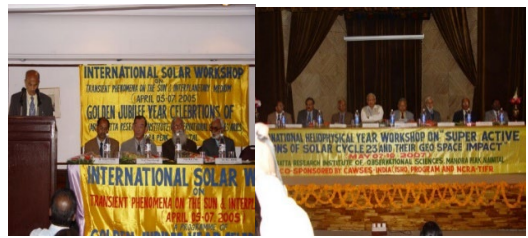

**Fig. 14** Left: Workshop on Transient Phenomena on the Sun and Interplanetary Medium and Right: workshop on Super Active Region of Solar Cycle 23 and their Geo-space Impact.

The left panel displays a snapshot of inaugural function of International Solar Workshop on *Transient Phenomena on the Sun and Interplanetary Medium* held during April 5-7, 2005. The dignitaries on the dais (from left to right) are author, Dr. Wahab Uddin, Prof. Takeo Kosugi (Japan), Prof. R.C. Pant and Dr. M.C. Pande. On the 50th anniversary of International Geo-physical Year (IGY) an International Helio-spheric Year (IHY) workshop on "*Super Active Region of Solar Cycle 23 and their Geo-space Impact*" was organized by ARIES during May 7-10, 2007 (right panel of Fig.14). Right panel shows a snapshot of inaugural function. Dignitaries on the dais (from left to right) are Dr. Wahab Uddin, Prof. N. Gopalswami (NASA, USA), Dr. P.B. Rao, Prof. C.P. Barthawal, Prof. A.P. Mitra, Dr. M.C. Pande, author and Prof. P.K. Manoharan. During the period of IGY (1957-58), India actively participated in IGY programs including successful optical tracking of Earth's artificial satellites with BN camera by the institute in collaborations with the SAO, USA. Dr. M.C. Pande's book on *Solar Physics* was released by Prof. A.P. Mitra. A ceremony was also organized to felicitate IGY Scientists for their significant contribution during IGY period. About 65 participants from different Institutes/Universities/Colleges from India and abroad and about 40 participants from ARIES attended the IHY Workshop. More than 40 oral and about 15 poster presentations were made. The proceedings of the Workshop were published in *Indian Journal of Radio and Space Physics* Vol 39 (2010).

## 3.3 Research in atmospheric science

Our earth is the only planet in our solar system where living beings exist. It is mainly because atmospheric conditions on the Earth provide air and suitable temperature for survival. However, during the last 300 years, human activities have disturbed the Earth's atmosphere that is leading to abnormal weather and climate changes. There have been observational evidences of an increase in concentrations of some of the greenhouse gases by 20% to 130% during past 1000 years. Our climate system is strongly influenced by the manner in which the incoming solar and outgoing longwave electromagnetic radiations interact with the aerosols, trace species and gases present mainly in the lower Earth atmosphere. These trace species, gases and aerosols, show large spatial and temporal variabilities that lead to significant uncertainty in quantifying their radiative forcing and hence making climate change predictions more difficult. This, however, becomes more complex on regional scale, e.g., processes in lower atmosphere become more complex in tropical regions due to more water vapor and intense solar radiation, leading to more active photochemistry. Also, over large region of Asia, aerosol clouds actually cause as much warming as greenhouse gases. In situ measurements of them from a number of places are, therefore, needed for understanding of climate and weather changes and deteriorating air quality on the Earth. Unfortunately, such measurements from a high-altitude site in central Gangetic Himalayan (CGH) region of Northern India were lacking till 2 decades ago and to fulfil this gap, ARIES started research in atmospheric science.

The Manora peak is located in the mountain region of the CGH at an aerial distance of ~ 2 km south-west from Nainital city which has no industrial activities. Also, population of the city is ~ 4 lakhs only. It is surrounded by sharply peaking mountains in the north and east direction and by mountains of altitudes < 1000 m in the south and west direction (Fig. 1). The nearby small cities having small-scale industrial activities are Haldwani and Rudrapur at altitude of 423 and 209 m respectively. They are located 20-40 km away from Manora peak in the south direction. The Delhi region, nearest city having major industrial activities, is located ~ 225 km away from Manora Peak in the west direction. Thus, both altitude and local topography of the Manora peak, are very well suited for the studies of fundamental questions related to climate change as well as to study the influence of intercontinental transport, mainly from Southern Europe and Northern Africa. The ISRO has, therefore, identified this site as one of the representative observational sites for Northern Indian region and has set up a complete environmental observatory in collaboration with ARIES.

### 3.3.1 History of atmospheric studies

During early years of the institute, airglow studies were carried out using photoelectric photometric observations of night airglow emissions (Sinha, 2005). Historically, surface-based observations of meteorology parameters, were started in July 1963. Manual observations of rainfall, relative humidity and temperature were initiated during that period. Later observations of winds, sunshine hours and pressure were also started. Two historical instruments operated till early 2000's at the institute, are shown in the two left panels of Fig. 15. They were used to record sunshine hours and pressure, based on strip chart recorder. These manual meteorological data, now digitized, are available for public. Now a days, observations of all meteorological parameters with a time resolution of 15-30 minutes are taken in digital mode with modern automatic weather station as shown in 3$^{rd}$ panel (from left) of Fig. 15.

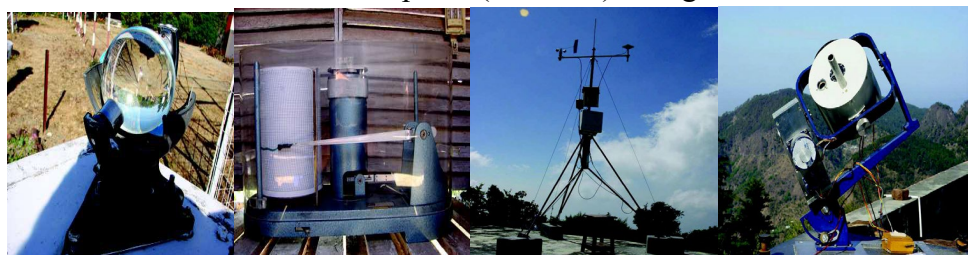

**Fig. 15** Historical instruments used for recording sunshine hours and pressure during 1960's.

Historical instruments used for recording sunshine hours and pressure during 1960's, are displayed in the two left panels. The extreme left panel shows a Campbell-Stokes sunshine recorder system which concentrates sunlight through a glass sphere onto a recording card placed at its focal point. The length of the burn trace left on the card represents the sunshine duration which is the length of time that the ground surface is irradiated by direct solar radiation. The 2nd panel from left display photo of a nautical barograph pressure measuring instrument. It records barometric pressure over time on a paper chart called barograms. A barograph has a metal cylinder which is connected with a pen arm. Pressure versus time is tracked down by the pen and is shown on the chart graphically. Third panel from left shows mechanical structure of an automatic weather station set up for digital recording of meteorological parameters with a time resolution of 15-30 minutes.  It consists of a weather-proof enclosure containing the data logger, rechargeable battery, telemetry and the meteorological sensors with an attached wind turbine mounted upon a mast. Muti-wavelength radiometer (MWR) set up is shown in the right most panel of the figure. This instrument was designed and developed by the Space Physics Laboratory, Thiruvananthapuram under ISRO – Geosphere Biosphere Program (GBP). It provides columnar total optical depths at ten narrow band wavelengths centered at 0.38, 0.40, 0.45, 0.50, 0.60, 0.65, 0.75, 0.85, 0.935 and 1.025 µm by making continuous spectral extinction measurements of directly transmitted solar radiation (Sagar et al. 2014 and references therein).

**Table 2**. Modern main atmospheric science instruments installed at Manora peak are listed.

| Observed parameters | Measurement Technique | Supplier of Instrument(s) | Year |
|---|---|---|---|
| Meteorology | Automatic Weather Station time resolution 15 -30 min | Campbell Scientific, Canada; Dynalab, India | 2004 |
| Aerosol optical depth (AOD) | Attenuation in the atmosphere | MWR from SPL and MICROTOPS from Solar Light Co., USA | 2002 |
| Aerosol number concentration | Optical particle counting | Grimm Aerosol Technik (1.108) | 2004 |
| Aerosol scattering | Scattering measured at 3 λ | Nephlometer (model 3563) | 2012 |
| Aerosol distribution | Probe-based imaging tool | APS, model 3321 | 2013 |
| Black Carbon | Optical Absorption | Magee Scientific (AE-42) | 2006 |
| NO and $NO_y$ | Chemiluminescence | Thermo (42 *i*) | 2006 |
| $SO_2$ | Pulsed Fluorescence | Thermo (43 *i*) | 2006 |
| CO | NDIR Spectroscopy | Thermo (48 *i*) | 2006 |
| Ozone | UV Absorption | Thermo (49 *i*) | 2006 |
| $CH_4$, $NMHC_s$, $CO_2$, $N_2O$, $SF_6$ | Weekly air sampling | Analysis using GC-FID | 2006 |

The instruments mentioned in Table 2 are used for in situ measurements of physical and chemical parameters of trace gas, aerosol and meteorology given in column 1 using the techniques/methods listed in column 2. The supplier of the instrument and the year when observations with the instrument started are listed in column 3 and 4 respectively. These observing facilities were installed by the ARIES mainly in collaboration with ISRO. Details of all these, modern state of the art observing facilities, are given by Sagar et al. (2014, 2015).

These observations are aimed for better understanding of complex processes (physical, chemical, and dynamical) that are governing our Earth's atmosphere in the CGH region.

Besides above, the observing facilities for the studies of very low frequency whistlers and AOD/flux measurements were also acquired by the ARIES. The balloon-borne observations of ozone and meteorological parameters such as relative humidity, temperature, and wind speed and wind direction are made every week since January 2011 using ECC-type ozone sonde (EN-SCI 2ZV7 ECC). The balloon ascent rate used to be ~5 m/s with burst altitude ~30 km and vertical resolution of ~10 m in the lower atmosphere. The sensors had a response time of 1–2 sec and accuracy of observed parameters was 5–10%.

The ARIES ST Radar (ASTRAD) system @ 206.5 MHz was developed and designed indigenously with funding from the DST. It is installed at the Manora peak campus of ARIES (Jaiswal et al., 2020). It is used for continuous measurements of 3-D vertical structure of the wind with very high temporal resolution. Such measurements made in all weather conditions offer an unparalleled opportunity to study not only gross features of the total wind field, but also small-scale, time-varying structures such as gravity waves, exchange process between stratosphere and troposphere, and turbulence through the middle atmosphere. Being the first and only such observational facility in the northern India, ASTRAD will boost atmospheric research in the CGH region. Fig. 16 shows photographs of foundation laying ceremony of ASTRAD building and the workshop entitled ***Radar Probing of the Atmosphere*** held during November 8 – 22, 2011. Renowned experts from different Universities/Institutes were invited to deliver lectures and invited talks to over more than 30 participants coming from different academic institutions of India.

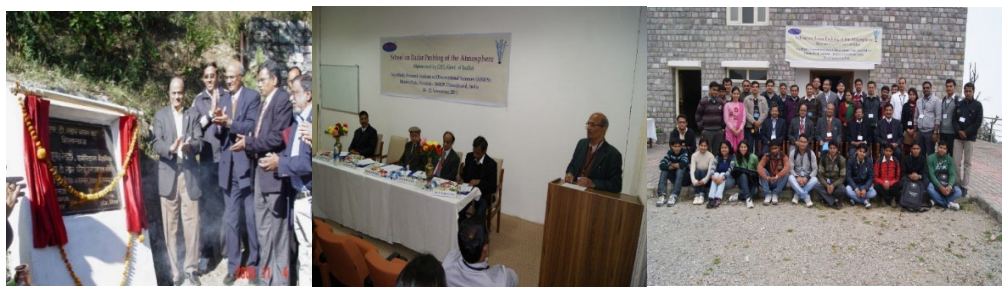

**Fig. 16** The foundation laying ceremony of ASTRAD building and the workshop entitled Radar Probing of the Atmosphere, 2009.

The left panel shows a snapshot photo of foundation stone laying ceremony of ST Radar building on November 4, 2009 by Prof. B. M. Reddy, Emeritus Scientist and Chairman of Project Management Committee, ST Radar. From left to right, Prof. B.M. Reddy, Mr. Chandra Prakash, author, Dr. P. Sanjeev Rao and Dr. G. Vishwanathan are seen in the photo. Central panel shows photo of an inaugural function of a DST sponsored school entitled: ***Radar Probing of the Atmosphere*** held during November 8 – 22, 2011 at ARIES Campus. Author is addressing the audience. Others seen in the photo, from left to right, are Dr Pitamber Pant, Prof. L. M. Singh Palani, Prof. B. M. Reddy and Dr. P. Sanjeev Rao. In the group photo (right most panel), taken in front of the ST Radar building, Dr. Manish Naja, Dr. Narendra Singh and Dr. D.V. Phanikumar, other staff members and PhD students from ARIES are seen with the participants (30) of the school.

Taking advantage of its geographical location, ARIES participated in a number of international campaigns. Only two major ones are mentioned here. First one, named RAWEX–GVAX (Regional Aerosols Warming Experiment – Ganges Valley Aerosol Experiment) was planned jointly under Indo-USA scientific collaboration as a multi-institutional, multi-instrument campaign. The first ARM (atmospheric radiation measurement) Mobile Facility (AMF-1) was deployed at the Manora peak from June 2011 to March 2012. Several aerosol and atmospheric parameters were measured extensively using state-of-the-art but commercially not available instruments mounted on the AMF-1. The RAWEX-GVAX carried out in situ measurements of a wide range of parameters like physical and optical properties of aerosols, meteorological parameters and boundary layer evolution etc.  These measurements were used to study the impact of aerosol on regional climate, contribution of Ganges Valley to aerosol plumes over Himalaya and to understand physical processes of aerosol–cloud interactions etc. An overview and major scientific findings of the experiment are published in a special issue of the Current Science published from Bengaluru (Moorthy et al., 2016). Second multi-institutional international Aircraft field campaign was conducted during year 2016-2017 under the European **StratoClim** project (Brunamonti et al., 2018). It investigated the effects of Asian summer monsoon anticyclone (ASMA) on meteorological system of the upper troposphere-lower stratosphere (UTLS) during summer. Observations taken during the campaign provided unprecedented insights into the UTLS thermal structure, the vertical distributions of water vapor, ozone and aerosols, cirrus cloud properties and interannual variability in the ASMA. All these ultimately contributed to our understanding of global climate change and the monsoon in the CGH region.

### 3.3.3   Significant atmospheric science contributions made by ARIES

Taking advantage of the geographical location of Manora Peak, ARIES has obtained valuable observations in the field of atmospheric science and published number of nationally and internationally recognised interesting scientific results (Sagar et al., 2014, 2015). Some of them are:
First ever and long duration (> 5 years) observations of aerosols over the CGH show that during wintertime, its pristine environment is dominated by fine aerosols (radius < 0.1 μm). This is comparable with those over Antarctic. In-contrast, summertime bigger aerosols loading is comparable to those over urban regions. It is due to dominating contribution of coarse aerosols (radius > 0.5 µm) from at-least two diverse and prominent sources of aerosols. Aerosols radiative forcing over the CGH is estimated to be very low (+4.9 W m$^{-2}$) in comparison to those over urban sites (+71 W m$^{-2}$). For the first time, an observational proof has also been provided that summertime transport of dust from Thar Desert is able to influence air-quality and radiation budget over the CGH region. The extended Lidar observations over Manora Peak substantiated presence of the elevated aerosol layers and clouds, which are important in the study of climate modelling.

Observations of trace gases (Ozone, CO, NONOy, CH4 and SO2) show that the photochemical ozone production is generally not significant over the CGH. Despite higher CO and NOy concentrations, ozone levels in this region are nearly similar to those at other global high-altitude sites.

Observations of ozone vertical distribution and meteorological parameters are made using balloon-borne sensors for the first time in the CGH region. Modelling of these observations suggest that regional photochemistry and biomass burning processes play controlling role in the lower troposphere, while, the middle-upper tropospheric variations are driven by dynamical processes including advection and stratospheric intrusion. All these ultimately contributed to our understanding of global climate change and the monsoon in the CGH region.

## 4   History of research publications and PhD awarded from the institute

A scientometric study indicates that the modern observational facilities of the ARIES provide opportunity not only for global collaborations but also for research publishable in high impact factor journals (Kumar, 2018).Table 3 shows number of research publications and PhD awarded from the institute during the period from 1954 to 2021. During entire period of 50 years under state government, the institute contributed to 36 PhD thesis and 782 research publications with an average of ~15.6 per year. About 2/3 of the total publications were in the reputed journals of the field. Number of publications in the field of atmospheric sciences, Galactic and extra-galactic and Solar physics were ~ 4 %, 64% and 32% respectively. After formation of ARIES, number of research publications and PhD thesis are 1103 and 60 respectively with an average of research publications of ~ 61 per year. These figures indicate ~ 4 times growth in research publications during last 18 years under DST in comparison to that of during 50 years under state governments. This could be due to the fact that after 2004, ARIES could attract some bright young scientists and engineers with excellent potential which even broadened the scope of ARIES by introducing *Atmospheric Science* as a new area of research. The number of scientific and engineering staff was, therefore, increased from about dozen in 2004 to 19, 27, 33 and 40 by March 31 of year 2005, 2007, 2009 and 2013 respectively. Numbers of the PhD students and post-doctoral fellows, project, visiting and adjunct scientists were also increased significantly from few in 2004 to over 50 in recent years. One of the main aim of ARIES formation was, thus, fulfilled i.e., to produce trained manpower and carry out front line research in the field of observational sciences.

**Table 3** Number of research publications and PhD thesis awarded during 1954 to 2021.

| Duration (years) | NT | PAM (%) | PAA (%) | PSP (%) | AY | NP |
|---|---|---|---|---|---|---|
| 1954 to 1970 | 62 | 18 | 56 | 26 | 4.1 | 2 |
| 1971 to 1980 | 154 | 5 | 62 | 33 | 15.4 | 8 |
| 1981 to 1990 | 265 | 1 | 61 | 38 | 26.5 | 12 |
| 1991 to 2003 | 301 | 3 | 69 | 28 | 23.2 | 14 |
| 2004 to 2013 | 515 | 14 | 71 | 15 | 51.5 | 32 |
| 2014 to 2021 | 588 | 25 | 68 | 7 | 73.5 | 28 |

Table 3 shows the number of research publications and PhD thesis awarded during 1954 to 2021. NT in 2nd column is the number of research publications during period listed in 1st column while PAM, PAA and PSP corresponding to 3rd, 4th and 5th columns respectively denote percentage of NT in the field of atmospheric sciences, Galactic and extra-galactic and Solar physics respectively. AY, average value of yearly publication and NP, number of PhD thesis awarded, are listed in the 6$^{th}$ and 7$^{th}$ column respectively.

In order to make best use of the funding received from DST, ARIES made micro-planning of all projects including the 3.6-m DOT costing over Rs. 150 Crores and ASTRAD costing over Rs 15 Crores for their speedy and timely implementation. A maximum of ~ Rs. 47. 5 Crores, ~3/4 only for the 3.6-m DOT project, was utilized by ARIES in the financial year 2008-2009. An average of over Rs. 28 Crores/year has been utilized by ARIES since 2004. This amount is similar to the total amount of Rs. 29.5 Crores utilized by the institution during entire period of 50 years under the state governments (Sinha, 2005). The DST provided much needed resources and environment for successful installation of number of front-line observational facilities as well as construction of infrastructural buildings at both Manora peak and Devasthal campuses of ARIES. All these as well as participation of ARIES in mega Indian astronomy projects like the Aditya, 2-m National large solar telescope and thirty-meter telescope indicate that formation of ARIES under DST augurs well for the growth of observational sciences globally.

## 5  Summary and conclusions

Idea of establishing an Observatory for the studies of celestial bodies was germinated about 7 decades ago jointly by Dr. Sampurnanand and Prof. A.N. Singh. Vision of Dr. M. K. Vainu

Bappu, a doyen and father of optical astronomy in India, was to make it an international institution for studies of celestial objects and events. For this, the institute installed nine optical telescopes ranging in aperture size from 25-cm gravity driven refractor to 3.6-m modern actively supported DOT and 4-m ILMT. The 104-cm Sampurnanand telescope has achieved an important historical mile stone of completing 50 golden years of its successful operation. Installation of optical telescopes at Devasthal, about 4 decades after installation of the 104-cm Sampurnanad telescope, has created important mile stones in the history of both ARIES and Indian science. Active optics system of the 3.6-m DOT compensates for small distortions in the shape of the 4.3 tonne primary mirror due to gravity or atmospheric aberrations. Optical and NIR observations taken with the 3.6-m DOT reveal that performance of the telescope is at par with other similar global telescopes. Till its operation in the year 2016, largest sized optical telescope in India was 2.34-m Vainu Bappu Telescope, inaugurated on January 6, 1986 by the then Prime Minister of India. Presently, the 1.04-m Sampurnanand telescope, 1.3-m DOFT and 3.6-m DOT are uses for observations of different types of Galactic and extra-galactic objects including follow up observations of GMRT, AstroSat and gravitational wave sources. Their geographical location has global importance particularly for the time domain astrophysical studies.

The location of the Manora Peak in the CGH region is globally recognised for studies related to climate change. The atmospheric science laboratory of ARIES provides *in situ* physical and chemical measurements of trace gases and aerosols for detailed study of the physical, chemical and dynamical processes in the lower atmosphere of the CGH region. Space archival observations combined with fast optical imaging observations of solar transients and TSE taken with observational facilities of ARIES are used to study physical processes responsible for various solar activities during different solar cycles and for the coronal heating of the Sun.

**Acknowledgements**

Author is grateful to the anonymous reviewer for providing constructive comments. Data and information given on the website of ARIES are extensively used. Help provided by Dr. Brijesh Kumar, Dr. Manish Naja and Dr. Amitesh Omar are thankfully acknowledged. During transfer of the institution from state to center, authorities of both Uttaranchal and DST provided valuable guidance. Thanks to the National Academy of Sciences, India, Prayagraj for the award of Honorary Scientist position; Alexander von Humboldt Foundation, Germany for the award of long-term group research linkage program and Director, IIA for providing infrastructural support during this work.